\pdfoutput=1
\documentclass[sigconf]{acmart}

\usepackage{microtype}

\copyrightyear{2022}
\acmYear{2022}
\setcopyright{acmlicensed}
\acmConference[SIGCSE 2022]{Proceedings of the 53rd ACM Technical Symposium on Computer Science Education V. 1}{March 3--5, 2022}{Providence, RI, USA}
\acmBooktitle{Proceedings of the 53rd ACM Technical Symposium on Computer Science Education V. 1 (SIGCSE 2022), March 3--5, 2022, Providence, RI, USA}
\acmPrice{15.00}
\acmDOI{10.1145/3478431.3499291}
\acmISBN{978-1-4503-9070-5/22/03}

\begin{document}
% \fancyhead{}

\title{CS Education for the Socially-Just Worlds We Need}
\subtitle{The Case for Justice-Centered Approaches to CS in Higher Education}

\author{Kevin Lin}
\orcid{0000-0001-9946-3635}
\affiliation{%
  \department{Paul G. Allen School of Computer Science \& Engineering}
  \institution{University of Washington}
  \streetaddress{185 E Stevens Way NE}
  \city{Seattle}
  \state{WA}
  \country{USA}
  \postcode{98195}
}
\email{kevinl@cs.uw.edu}

\begin{abstract}
Justice-centered approaches to equitable computer science (CS) education frame CS learning as a means for advancing peace, antiracism, and social justice rather than war, empire, and corporations. However, most research in justice-centered approaches in CS education focus on K--12 learning environments. In this position paper, we review justice-centered approaches to CS education, problematize the lack of justice-centered approaches to CS in higher education in particular, and describe a justice-centered approach for undergraduate Data Structures and Algorithms. Our approach emphasizes three components: (1) ethics: critiques the sociopolitical values of data structure and algorithm design as well as the underlying logics of dominant computing culture; (2) identity: draws on culturally responsive-sustaining pedagogies to emphasize student identity as rooted in resistance to the dominant computing culture; and (3) political vision: ensures the rightful presence of political struggles by reauthoring rights to frame CS learning as a force for social justice. Through a case study of this \emph{Critical Comparative Data Structures and Algorithms} pedagogy, we argue that justice-centered approaches to higher CS education can help all computing students not only learn about the ethical implications of nominally technical concepts, but also develop greater respect for diverse epistemologies, cultures, and experiences surrounding computing that are essential to creating the socially-just worlds we need.
\end{abstract}

\begin{CCSXML}
<ccs2012>
  <concept>
    <concept_id>10003456.10003457.10003527</concept_id>
    <concept_desc>Social and professional topics~Computing education</concept_desc>
    <concept_significance>500</concept_significance>
  </concept>
</ccs2012>
\end{CCSXML}
\ccsdesc[500]{Social and professional topics~Computing education}

\keywords{%
    critical pedagogy;
    cultural competence;
    culturally responsive pedagogy;
    data structures;
    ethics;
    identity;
    political vision;
    social justice
}

\maketitle

\section{Introduction}

Dominant approaches to CS education frame ``equity as inclusion'' \cite{CalabreseBarton2020BeyondLearning}, emphasizing (for example) \emph{capacity}, \emph{access}, and \emph{participation} in CS education for all students \cite{Fletcher2021CAPE}. This assumes that all students want to learn CS the way we currently teach it. ``However, are we satisfied with everyone learning to code, if the end game is to produce (admittedly more `diverse') coders who will primarily work to ensure the continued profitability of capitalist start-ups and technology giants?'' \cite{Costanza-Chock2020DesignNeed}. Are the more `diverse' students we wish to recruit and retain satisfied with this vision for CS?

Beyond \emph{capacity}, \emph{access}, and \emph{participation}, \citeauthor{Fletcher2021CAPE} describe student \emph{experience} as the fourth component of the CAPE framework for assessing equity in CS education. While student performance measures such as course grades provide one means of assessing equity in outcomes, they argue that ``truly equitable experiences must go beyond these simple outcome measures'' in order to ``create an environment where all students feel they belong, instruction is inclusive, and diverse perspectives are valued explicitly'' \cite{Fletcher2021CAPE}. Proponents of critical CS education \cite{Ko2020ItEducation, Ryoo2019PedagogyAll} argue that teaching computer science content knowledge alone is not enough ``because they do not consider the perspective of the group being served'' \cite{HubbardCheuoua2021ConfrontingTheory}. Dominant approaches to CS education marginalize diverse students' identities and political values by positioning CS education as a force for reproducing social inequity---more interested in maximizing efficiency and profit than doing good \cite{CalabreseBarton2020BeyondLearning, Vakil2018EthicsEducation, Ryoo2019PedagogyAll, Ryoo2020TakeSchools, Davis2021CulturallyFramework, Vakil2020IveEducation, Shah2020RacialClassrooms, Rankin2020TheComputing, Vakil2019ExploringClassroom}.

Informed by critical theory, justice-centered approaches to CS education complicate and challenge prior research efforts that ``explain Black (and other minoritized) students' motivations about what to learn or not to learn as tied to their perceptions of what is `geeky' (or inversely what is `cool'),'' instead emphasizing ``more complex dynamics underlying student resistance or interest'' toward learning CS such as students' identities and political values \cite{Vakil2020IveEducation}. Justice-centered approaches are currently emphasized in K--12 CS education through curricula such as Exploring Computer Science \cite{Ryoo2019PedagogyAll}, but they are relatively absent in higher CS education research and practice. In this position paper, we argue for justice-centered approaches to CS in higher education. Section~\ref{sec:justice-centered-cs-education} reviews the literature on justice-centered approaches that center ethics, identity, and political vision. Section~\ref{sec:why-justice-in-higher-cs-education} makes the case for attention to justice-centered approaches in higher CS education. Section~\ref{sec:data-structures-and-algorithms} problematizes dominant approaches to teaching undergraduate Data Structures and Algorithms (CS2). Section~\ref{sec:critical-comparative-data-structures-and-algorithms} proposes a novel, justice-centered approach for CS2 by applying critical pedagogy. Our justice-centered CS2 presents a case study of how a course that has traditionally marginalized all three of ethics, identity, and political vision can be reauthored to center justice.

\section{Justice-Centered CS Education}
\label{sec:justice-centered-cs-education}

\citeauthor{Vakil2018EthicsEducation} defines justice-centered approaches to CS education as attending to three features: the content of curriculum (centering ethics), the design of learning environments (centering identity), and the politics and purposes of CS education reform \cite{Vakil2018EthicsEducation}.

\subsection{Ethics in the computing curriculum}

Recent work in undergraduate computing ethics include standalone ethics courses \cite{Reich2020TeachingEthics, Ferreira2021DeepScience}; integrated ethics across the curriculum \cite{Grosz2019EmbeddedEthiCS, Cohen2021ACurriculum}; and integrated ethics modules or lessons in courses such as machine learning \cite{Saltz2019IntegratingCourses}, human-centered computing \cite{Skirpan2018EthicsContext}, and introductory CS \cite{Fiesler2021IntegratingClasses, Doore2020AssignmentsTechnology}. \citeauthor{Fiesler2020WhatEthics} analyzed 115 syllabi from tech ethics courses and found that ``many topics within tech ethics are high level and conceptual when it comes to the impact of technology on society---e.g., how human decisions are built into code, how technology can reproduce and augment existing social inequalities, how data is created by and directly impacts people, and how choices made at both the level of companies and in small bits of code combine to create large-scale social consequences'' \cite{Fiesler2020WhatEthics}.

Critical approaches to tech ethics extend these ideas by emphasizing computing's political power to reshape social structures and hierarchies in the eye of the designer \cite{Ko2020ItEducation, Vakil2018EthicsEducation, Vakil2020IveEducation, Ferreira2021DeepScience, Raji2021YouUs, Moore2020TowardsScience, Ryoo2020TakeSchools, Vakil2019ItsPower}. Although \citeauthor{Fiesler2020WhatEthics} argue that tech ethics ``could be part of every computing course'' \cite{Fiesler2020WhatEthics}, the inclusion of tech ethics is challenged by the hierarchy of knowledge in computing that prioritizes ``technical'' skills over ``social'' skills \cite{Raji2021YouUs}. An analysis of 200 ``technical'' artificial intelligence and machine learning courses by \citeauthor{Saltz2019IntegratingCourses} revealed only 12\% of courses included some mention of ethics; and of these 12\% of courses, ethics-related topics were relegated to the last two classes in the schedule and, in one course, left as a discussion topic only ``if time allows'' \cite{Garrett2020MoreEducation}.

The exclusion of ethics in computing courses produce a kind of epistemic violence where dominant, ``technical'' ways of knowing CS are reinforced while marginalized, ``social'' ways of knowing CS are diminished. \citeauthor{Malazita2019InfrastructuresSubjects} observe this tension embedded in the emphasis on abstraction: the fundamental concept that some details or information are more relevant to a computational solution than others. The absence of ethics combined with the emphasis on abstraction defines CS education as not just apolitical (disregarding political or ethical values) but rather \emph{anti-political}: computing is framed as a means of ``solving'' complex social problems by defining abstractions to manage complexity \cite{Malazita2019InfrastructuresSubjects}.

But critical educators argue that this framing is problematic: ``quick ethics fixes, like modules largely developed for and within computer science, are not a sufficient intervention to actually teach CS students of how ethical challenges get resolved in real world contexts'' \cite{Raji2021YouUs}. Rather, more interdisciplinary efforts are needed to engage students with diverse stakeholders and collaborators \cite{Raji2021YouUs, Vakil2019ItsPower}. In \emph{Design Justice}, \citeauthor{Costanza-Chock2020DesignNeed} calls on us to ``seek more than `freedom from bias.' For example, feminist and antiracist currents within science and technology studies have gone beyond a bias frame to unpack the ways that intersecting forms of oppression, including patriarchy, white supremacy, ableism, and capitalism, are constantly hard-coded into designed objects, platforms, and systems'' \cite{Costanza-Chock2020DesignNeed}. Justice-centered approaches must teach students about the relationships between computing, power, and identity.

\subsection{Identity in the learning environment}

Students have not only social identities such as race, gender, or ethnicity, but also disciplinary identities that represent what they might be able to do with computer science \cite{Vakil2020IveEducation}. Although data on social identity is often prominent in CS education research (e.g. racial demographics), empirical or theoretical data on disciplinary identity is often lacking \cite{Vakil2018EthicsEducation} despite the importance of disciplinary identity in shaping students' sensemaking around the values of CS \cite{Vakil2020IveEducation}. For marginalized students who experience firsthand the inequities in the social structures of schooling, their disciplinary identity is inextricably linked to their political identity and their commitment to issues of power and social justice \cite{Vakil2018EthicsEducation, Vakil2020IveEducation, CalabreseBarton2020BeyondLearning}. Yet dominant approaches to CS education that frame learning as anti-political (where only the ``technical'' ideas count) emphasize that CS has no space for students' political identities.

Disciplinary identity is not only political, but also intersectional. Just as \citeauthor{Costanza-Chock2020DesignNeed} emphasizes the intersecting oppressions hard-coded in designed objects, students' disciplinary identities are also shaped by the ``power dynamics that exemplifies racism, sexism, socioeconomic status, homophobia, ableism, xenophobia, etc.'' \cite{Rankin2021RealEducation}. To create more inclusive and anti-oppressive learning environments, \citeauthor{Washington2020WhenEnough} argues for cultural competence in computing (3C): greater awareness, attitudes, knowledge, and skills toward working effectively in cross-cultural situations \cite{Washington2020WhenEnough}. To realize this, \citeauthor{Davis2021CulturallyFramework} define a framework for culturally responsive-sustaining CS pedagogy to ensure that ``students' interests, identities and cultures are embraced and validated, students develop knowledge of computing content and its utility in the world, strong CS identities are developed, and students engage in larger socio-political critiques about technology's purpose, potential, and impact'' \cite{Davis2021CulturallyFramework}.

\subsection{Political vision for CS education}

Given the importance of political identity toward students' disciplinary identity and their understanding of tech ethics, justice-centered approaches engage political identity by ``collectively, clearly, and unequivocally articulat[ing] a political vision for CS learning anchored in principles of peace, antiracism, and justice'' that ``challenges the corporate technology sector on moral, epistemological, and political grounds'' \cite{Vakil2018EthicsEducation}. Justice-centered approaches seek to develop students' sociopolitical consciousness: ``the recognition and desire to act upon societal inequities'' through computing \cite{Madkins2019IlluminatingInstruction}.

Integrating disciplinary identity and sociopolitical consciousness, \citeauthor{CalabreseBarton2020BeyondLearning} describe a framework of \emph{rightful presence} ``towards \emph{making present} the intersections of contemporary (in)justices, while orienting towards new, just social futures'' \cite{CalabreseBarton2020BeyondLearning}. \citeauthor{Ko2020ItEducation} argue for making injustices visible in CS education; rightful presence does so by foregrounding social and political narratives as a core pedagogical practice. CS education's commitment to justice in the classroom and beyond can be enacted through political activism in the classroom, such as ``calls to action, practitioner reflections, legislative engagement, and direct action'' \cite{Moore2020TowardsScience}. This political activism moves beyond narrow ethical critiques or modules \cite{Raji2021YouUs} and supports student agency to create the socially-just worlds we need \cite{Costanza-Chock2020DesignNeed}. By articulating a just political vision for CS education, ethics and identity ``do not need to be `included' in the curriculum'' because they will already be the center of inquiry \cite{Vakil2018EthicsEducation}.

\section{Why Justice in Higher CS Education}
\label{sec:why-justice-in-higher-cs-education}

Recent research calls attention to the limitations of dominant and epistemologically-exclusionary approaches to ethics in higher CS education \cite{Raji2021YouUs, Malazita2019InfrastructuresSubjects, Cohen2021ACurriculum}. These critiques reveal problems arising from the lack of attention to identity and political vision: CS education that ultimately produces students whose anti-political values perpetuate injustice \cite{Malazita2019InfrastructuresSubjects, Costanza-Chock2020DesignNeed, CalabreseBarton2020BeyondLearning}; educators who feel the need to avoid rather than address ``political overtones'' in ethics education \cite{Cohen2021ACurriculum}; internship experiences that marginalize and oppress Black women in computing \cite{Rankin2021RealEducation}; and ``well-intentioned'' but culturally-blind CS faculty \cite{Malazita2019InfrastructuresSubjects, Washington2020WhenEnough} whose resistance to critical interventions and identity-centered instruction marginalizes students' political identities \cite{Vakil2018EthicsEducation, Vakil2019ItsPower, Vakil2020IveEducation, Ryoo2020TakeSchools, Philip2021TheoriesComputing}. \citeauthor{Ko2020ItEducation} urge CS educators to make injustices visible ``through the problems we focus on in our classrooms; through who we choose to teach; in how we shape students' career choices; and in how we conceptualize computing to journalists, social scientists, and society'' because ``[t]he world has critical questions about computing and it is time we started teaching more critical answers'' \cite{Ko2020ItEducation}. But much of the research on justice-centered approaches to CS education focuses on K--12 learning environments \cite{Vakil2018EthicsEducation, Vakil2019ExploringClassroom, CalabreseBarton2020BeyondLearning, Ryoo2019PedagogyAll, Ryoo2020TakeSchools, Madkins2019IlluminatingInstruction, Shah2020RacialClassrooms}; much less research focuses on higher education. Ethics without identity or political vision is problematic.

We need justice-centered approaches to higher CS education to support justice-centered K--12 CS education; to help programs make progress toward diversity, equity, inclusion, and access (DEIA) goals and broaden participation in computing (BPC); and to develop student interest toward creating more socially-just worlds.

\subsection{Support justice-centered K--12 CS education}

Justice-centered approaches to higher CS education support the parallel and ongoing efforts in K--12 CS education across curricula such as Exploring Computer Science \cite{Ryoo2019PedagogyAll}, books such as \emph{Critically Conscious Computing: Methods for Secondary Education},\footnote{\url{https://criticallyconsciouscomputing.org/}} and teacher education programs such as the University of Washington's STEP CS.\footnote{\url{https://criticalcsed.org/program/}} The current dominant approaches to higher CS education place critically-conscious secondary educators in a difficult (and potentially untenable) position. Given the need to \emph{make present} contemporary injustices, critical secondary educators must make clear to their students that what awaits them in higher CS education is an epistemologically-exclusive and anti-political experience \cite{Vakil2020IveEducation, Malazita2019InfrastructuresSubjects}---one that often rejects ethics, marginalizes students' social and political identities, and envisions computing as a force for corporate profit. Higher CS education risks not only undoing efforts in K--12 CS education, but also subjecting undergraduate CS students to epistemic, material, and physical harm \cite{Rankin2021RealEducation, Vakil2018EthicsEducation, Philip2021TheoriesComputing, Ko2020ItEducation}.

By making present these injustices, students are less likely to put up with oppressive education systems. As justice-centered approaches become increasingly common, critically-conscious students unsatisfied with dominant approaches to higher CS education know to vote with their feet and enroll in programs that support their CS identity. Institutions unresponsive to justice-centered approaches risk backsliding on recruiting and retaining diverse students by failing to center marginalized students' identities and political values in computing.

\subsection{Make progress toward DEIA goals and BPC}

Justice-centered approaches to higher CS education can help undergraduate computing programs realize diversity, equity, inclusion, and access (DEIA) goals and broaden participation in computing (BPC). These goals are not only initiated within institutions of higher CS education, but also mandated by funding agencies for certain research grants: the National Science Foundation Computer and Information Science and Engineering directorate, for example, recently began requiring principal investigators of proposals submitted to selected programs to include a plan for broadening participation in computing at the time of award.\footnote{\url{https://www.nsf.gov/cise/bpc/}} \citeauthor{Fletcher2021CAPE} describe ``CAPE: A Framework for Assessing Equity throughout the Computer Science Education Ecosystem'' that addresses \emph{capacity for}, \emph{access to}, \emph{participation in}, and \emph{experience of} equitable CS education \cite{Fletcher2021CAPE}. Justice-centered approaches to CS education directly affects student experience in the way they ``explicitly address issues of equity'' and help ``all students feel included and accepted'' \cite{Fletcher2021CAPE}.

Attention to all three features of justice-centered approaches---ethics, identity, and political vision---can help higher CS education move ``beyond equity as inclusion'' \cite{CalabreseBarton2020BeyondLearning} and address disparities in student participation and experience by ensuring the rightful presence of marginalized students' interests in learning CS \cite{CalabreseBarton2020BeyondLearning, Ryoo2020TakeSchools}. Justice-centered approaches can improve DEIA in programs and support BPC efforts by centering the values, experiences, and purposes of marginalized students through emphasis on computing's social responsibility \cite{Cohen2021ACurriculum, Ko2020ItEducation, Vakil2018EthicsEducation}, the disparate experiences of students with dominant versus marginalized identities \cite{Rankin2020TheComputing, Rankin2021RealEducation, Washington2020WhenEnough, Shah2020RacialClassrooms}, and the political vision of computing toward realizing more socially-just futures \cite{Costanza-Chock2020DesignNeed, Vakil2018EthicsEducation, Ko2020ItEducation}.

\subsection{Reauthor CS for more just futures}

Despite differences in curricula between institutions, higher CS education shares common authors: the 1960s and 1970s academic computer scientists whose dominant, European scientific values \cite{Aikenhead2007IndigenousRevisited} inspired a vision for computing that centered cognition and mathematics \cite{Vakil2018EthicsEducation, Kafai2019FromDialogue, Raji2021YouUs} to the exclusion and marginalization of identity and political vision \cite{Vakil2018EthicsEducation, Vakil2019ItsPower, Vakil2020IveEducation, Ryoo2020TakeSchools, Philip2021TheoriesComputing}. Justice-centered approaches to higher CS education enable teachers to engage students in a process of \emph{reauthoring rights} to value more diverse epistemologies rather than expecting assimilation to dominant epistemologies \cite{CalabreseBarton2020BeyondLearning}.

The implications of \emph{reauthoring} extend beyond the classroom because they involve communication and learning not only with students, but also with teachers as they navigate tensions between values and purposes for CS education. Reauthoring considers students' cultural values, experiences, and ways of knowing as ``integral to disciplinary learning'' \cite{CalabreseBarton2020BeyondLearning}. To address today's critical questions surrounding social computation, dominant approaches that emphasize an epistemological wall between the ``technical'' and the ``social'' are not equipped to provide critical answers \cite{Ko2020ItEducation}. Reauthoring rights through a process of political struggle and critical engagement---involving both students and teachers---can move computing as a discipline toward more critical perspectives that better appreciate and understand the sociopolitical implications of computation for all. In doing so, we develop student interest toward creating more socially-just worlds.

\section{Data Structures and Algorithms}
\label{sec:data-structures-and-algorithms}

In this section, we propose a justice-centered approach to teaching undergraduate Data Structures and Algorithms, or ``CS2'' at many institutions. While CS2 is a broad label representing the second course in computer science, many CS2 courses emphasize the design, implementation, and application of data structures and algorithms. Through conversation with experienced instructors, \citeauthor{Porter2018DevelopingCS2} identified ``two largely disjoint courses that are referred to in the CS education community as CS2''---Basic Data Structures and Advanced Data Structures \cite{Porter2018DevelopingCS2}. They argue that, ``At the end of a course on Basic Data Structures, students should be able to:
\begin{enumerate}
  \item Analyze runtime efficiency of algorithms related to data structure design.
  \item Select appropriate abstract data types for use in a given application.
  \item Compare data structure tradeoffs to select the appropriate implementation for an abstract data type.
  \item Design and modify data structures capable of insertion, deletion, search, and related operations.
  \item Trace through and predict the behavior of algorithms (including code) designed to implement data structure operations.
  \item Identify and remedy flaws in a data structure implementation that may cause its behavior to differ from the intended design.
\end{enumerate}
Advanced Data Structures extend these learning objectives to emphasize ``topics that rely on earlier data structures (e.g., balanced trees rely on BSTs, heaps rely on arrays)'' as well as graph representations and graph algorithms \cite{Porter2018DevelopingCS2}.

\subsection{Dominant approaches to teaching CS2}

Dominant approaches to teaching CS2 center the development of cognitive skills toward relatively standardized data structures and algorithms content knowledge: CS learning is ultimately about increasing ``individual comprehension of CS concepts and competent programming performance'' \cite{Kafai2019FromDialogue}. Although there is much to critique about dominant approaches, this \emph{cognitive framing} for CS learning does not necessarily imply that learning is decontextualized or irrelevant to students. The persistent popularity of the ``Nifty Assignments'' session at the annual ACM Technical Symposium on Computer Science Education (SIGCSE) is a testament to CS1 and CS2 educators' desire to design and adapt assignments that connect to students' diverse interests for learning CS.

However, without justice-centered approaches, CS2 instructors risk reproducing present-day oppressions. By teaching data structures as implementations for abstract data types, undergraduate computing programs emphasize the dominant programming practice that uses abstract data types to free ``a programmer from concern about irrelevant details in his use of data abstractions'' \cite{Liskov1974ProgrammingTypes}---the kind of content knowledge reinforcing the \emph{infrastructures of abstraction} that produce anti-political values \cite{Malazita2019InfrastructuresSubjects}. By centering runtime or space complexity analysis \cite{Porter2018DevelopingCS2} as the primary (and sometimes only) method for evaluating data structure and algorithm tradeoffs or design decisions, undergraduate computing programs emphasize efficiency as the primary metric for determining the quality of a computational solution. A more critical reading of the CS2 learning goals reveals the limits of cognitive approaches that frame algorithm design and implementation as ``a means of realizing a specification or abstract data type without critically questioning the design of the abstraction'' \cite{Lin2021DoAnalysis, Costanza-Chock2020DesignNeed, Malazita2019InfrastructuresSubjects}. CS2's content knowledge emphasis on implementations over abstractions presents a unique challenge for critically-conscious and justice-centered approaches. Sociopolitical values are encoded in the design of abstractions \cite{Lin2021DoAnalysis, Costanza-Chock2020DesignNeed}, but CS2's emphasis on implementations avoids discussion of the difficult social contexts and instead focuses on purely cognitive and mathematical analyses.

Justice-centered approaches are not necessarily mutually exclusive to the dominant cognitive approach \cite{Kafai2019FromDialogue}. In the context of CS2, undergraduate Data Structures and Algorithms content knowledge is uniquely powerful: completion of a ``Data Structures and Algorithms'' course offers a kind of limited certification to corporations that a student meets some of the qualifications for software engineering jobs. Completion of CS2 offers students easily-realizable power afforded by social mobility because data structures and algorithms content knowledge can prepare students for internships and full-time work in computing industries. The financial opportunities afforded by access to high-paying jobs in computing research and industry offers significant value to oppressed students who might otherwise have few options to generate wealth, sustain their communities, and escape poverty.

Where dominant approaches fail is in their inattention to the participation and experience of marginalized students in computing. Dominant approaches to CS2 risk exacerbating inequity in computing by creating and sustaining wealth for dominant students and corporations overwhelmingly benefiting from higher CS education's production of anti-political programmers \cite{Vakil2018EthicsEducation, Costanza-Chock2020DesignNeed, Malazita2019InfrastructuresSubjects}. After all, anti-political programmers are good for maximizing profit: they're easy to control and unlikely to resist because they see their design and engineering work as separate from the world. A justice-centered approach to Data Structures and Algorithms not only equips students with cognitive skills that unlock high-paying computing jobs, but also teaches students how they might navigate the tension around ``selling out'' their political commitments just to be ``a part of a huge unfeeling oppressive corporation that makes you money sure, but never does something good'' \cite{Vakil2020IveEducation}.

\section{Critical Comparative\\Data Structures and Algorithms}
\label{sec:critical-comparative-data-structures-and-algorithms}

\emph{Critical Comparative Data Structures and Algorithms} (CCDSA) is a novel, justice-centered approach for teaching undergraduate Advanced Data Structures.
\begin{description}
  \item[Ethics] Critiques sociopolitical values of data structure and algorithm design and dominant computing epistemologies that approach social good without design justice.
  \item[Identity] Centers students in culturally responsive-sustaining pedagogies to resist dominant computing culture and value Indigenous ways of living in nature.
  \item[Political vision] Ensures the rightful presence of political struggles through reauthoring rights and emphasizes the political power of computing  as a force for social justice in contrast to dominant narratives around corporate profit and hegemony.
\end{description}
The approach relies on \emph{critical comparison} as the primary method of inquiry for centering ethics, identity, and political vision. The method draws on traditions in critical pedagogy wherein structural critiques are foregrounded in students' education and injustices are made present. \citeauthor{Kafai2019FromDialogue} coined a more researcher-centered view of the critical comparison approach: \emph{theory dialogue}, which is designed to engage the diversity of cognitive, situated, and critical framings for CS education by emphasizing ``understanding of key computational concepts, practices, and perspectives'' (cognitive framing); ``stress[ing] personal creative expression and social engagement'' (situated framing); and respecting ``the values, practices, and infrastructure underlying computation as part of a broader goal of education for justice'' (critical framing) \cite{Kafai2019FromDialogue}.

CCDSA does not strictly necessitate removing traditional, cognitive content from CS2. Instead, the critical comparison method enables learning of dominant knowledge to be framed from several diverse epistemological perspectives---the situated and critical framings often marginalized by dominant approaches. Although foregrounding the critical framing does not completely erase the tensions around ``selling out,'' it offers students ``a new and exciting possibility to \emph{be} political while engaging in creating technology within the context of [a] CS class'' \cite{Vakil2020IveEducation}.

\subsection{Ethics via epistemological comparison}

The epistemological values of dominant approaches result in ``ethical and social interventions in CS education becom[ing] framed as valuable in application-centered classes, like data visualization or applied machine learning, but not in `core' technical classes'' like CS2 \cite{Malazita2019InfrastructuresSubjects}. CCDSA proposes countering the dominant narrative by incorporating ethics as a type of algorithm analysis on equal epistemological footing as runtime or space complexity analysis.

CCDSA engages ethics in CS2 with an affordance analysis of data structures and algorithms \cite{Lin2021DoAnalysis}: a more critical algorithm analysis that draws on critical methods from science and technology studies, philosophy of technology, and human-computer interaction in order to evaluate the political consequences of data structures and algorithms in social contexts and applications \cite{Lin2021DoAnalysis}. Unlike dominant approaches to ``algorithm analysis'' that emphasize the internal implementation of data structures and algorithms \cite{Porter2018DevelopingCS2}, affordance analysis emphasizes the external interface of data structures and algorithms as they are applied in real-world applications. While dominant approaches to teaching abstraction risk producing ``CS students as knowers who organize the world through excision'' and abstraction \cite{Malazita2019InfrastructuresSubjects}, affordance analysis specifically problematizes the affordances of abstract data types (e.g. priority queues) as they are applied to problems (e.g. content moderation) by emphasizing how the design of an abstraction encodes affordances that can have political values \cite{Lin2021DoAnalysis}. While comparison between data structures and algorithms on the basis of efficiency are important, CCDSA emphasizes \emph{critical comparison between abstractions on the basis of their sociopolitical implications}.

However, CCDSA is not satisfied with affordance analysis alone. Although affordance analysis makes space for sociotechnical critique of the design of computational solutions, ``it is inherently limited to the algorithmic components of a sociotechnical system'' and provides less instruction toward redressing harmful design values \cite{Lin2021DoAnalysis}, engaging epistemological limits \cite{Malazita2019InfrastructuresSubjects}, or reauthoring the political purposes of CS \cite{CalabreseBarton2020BeyondLearning, Vakil2020IveEducation}. \citeauthor{Costanza-Chock2020DesignNeed} argues for practices that move beyond the ``universalizing assumptions behind affordance theory'' and instead attend to design justice that pushes students to ``think more critically about software, technology, and design [\textellipsis] in service of human liberation and ecological sustainability'' \cite{Costanza-Chock2020DesignNeed}. Echoing the centrality of ethics, identity, and political vision in justice-centered CS education, design justice emphasizes the design practices, design narratives, design sites, and design pedagogies that create social conditions. Ethical analysis is not just limited to the harmful computer technology itself, but also all of the processes, infrastructures, and cultures that informed and enabled its design, implementation, and deployment \cite{Costanza-Chock2020DesignNeed}.

CCDSA engages design justice to move beyond affordance analysis toward \emph{critical comparison of computing epistemologies and design narratives}. Dominant approaches that center the software designers/engineers are compared to justice-centered approaches that position community members as the rightful designers. Design justice highlights the need to reauthor relationships between ``designer'' and ``user'' in order to give power back to people, rather than corporations or governments that often design technologies with little structural input from the people that they would most greatly affect. ``Design justice is interested in telling stories that amplify, lift up, and make visible existing community-based design solutions, practices, and practitioners'' \cite{Costanza-Chock2020DesignNeed}. Design justice often manifests in justice-centered approaches to K--12 CS education through projects enable students to design for their own communities, needs, and interests \cite{Ryoo2020TakeSchools, Vakil2020IveEducation, Kafai2019FromDialogue, Davis2021CulturallyFramework}. In CCDSA, design justice counters the dominant approach that emphasizes software design as elite, private, and exclusionary; and instead centers community goals and values.

\subsection{Identity via cultural comparison}

CCDSA implements 5 of the 6 core components for culturally responsive-sustaining CS pedagogy \cite{Davis2021CulturallyFramework} to center student identity. Although the framework emphasizes all three of ethics, identity, and political vision, in this section we'll focus on how it specifically enables \emph{critical comparison} of cultural approaches to computation.

\subsubsection{Acknowledge racism in CS and enact antiracist practices}

The instructor leads fireside chats with teaching assistants (TAs) or students to explore their social identities (e.g. race, gender, ethnicity), their power and privileges, and how lived experiences have shaped their worldviews. By calling-out sites of inequity in the CS classroom and beyond, the teaching staff makes a commitment to dismantling structural oppression and decenter whiteness \cite{Davis2021CulturallyFramework, Rankin2021RealEducation, Shah2020RacialClassrooms}. \emph{Critical comparison} is made between dominant European scientific culture and Indigenous ways of living in nature \cite{Aikenhead2007IndigenousRevisited} to engage the limits of European scientific epistemology and knowledge.\footnote{\url{https://courses.cs.washington.edu/courses/cse373/22wi/}}

\subsubsection{Create inclusive and equitable classroom cultures}

Antiracist practices are reinforced through course structure, assignments, and policies. Instruction and assessment applies universal design for learning to meet students' diverse means of engagement, representation, and action/expression \cite{Burgstahler2011UniversalEducation} by valuing many reasons to learn data structures and algorithms. A specifications-based grading system enables creative assessments (e.g. self-reflections and student-submitted video explanations) without assigning a demeaning, trivial percentage weight. The creative assessments give students full opportunity to collaborate on most of the work in the course. Students are not only encouraged to collaborate but also taught how to recognize, confront, and dispel stereotypes and power imbalances that occur during teamwork \cite{Shah2020RacialClassrooms}.

\subsubsection{Pedagogy and curriculum are rigorous, relevant, and encourage sociopolitical critiques}

CCDSA recognizes the power afforded to students by the cognitive framing (because CS2 offers social mobility) but problematizes the applications of data structures and algorithms toward social problems embedded in complex historical and sociopolitical context. Dominant approaches emphasize programming implementation and asymptotic analysis, which admit little creativity and often leave students ``subservient to the micro-demands of the autograder'' \cite{Malazita2019InfrastructuresSubjects}. \emph{Critical comparison} engages students beyond programming and asymptotic analysis toward sociopolitical critiques of technology in relation to people, places, values, and hierarchies.\footnote{\url{https://github.com/kevinlin1/huskymaps}}

\subsubsection{Student voice, agency, and self-determination are prioritized in CS classrooms}

Evidence-based pedagogies such as POGIL \cite{Kussmaul2012ProcessScience} offer a more structured approach to teamwork in order to ensure equity and teach students process skills such as communication and team management. POGIL classroom instruction is designed to amplify student voices rather than instructor voices. In particular, the instructor never presents ideas on their own. Instead, they listen to student teamwork on problems during class and amplify ideas generated during teamwork. Teaching assistants are specifically taught to resist the dominant pedagogical culture of teacher-centered instruction. Problem sets previously handed out during recitations are integrated into the main class and recitations are redesigned around self-reflection, team reflection, and whole-recitation reflection. Cognitive ideas, process skills, and emotional experiences are foregrounded through discussion with peers during recitations.

\subsubsection{Family and community cultural assets are incorporated into CS classrooms}

Community and cultural engagement beyond the scope of individual student experiences are not easily engaged by the dominant approaches to higher STEM education---particularly in large, research-oriented universities where impact is more largely determined by national prestige than service to the local community. Future work could examine the ways that higher CS education can partner undergraduate students with local community organizations in ways that are synergistic with the CS2 curriculum.

\subsubsection{Diverse professionals and role models provide exposure to a range of CS/tech careers}

The teaching staff are often the most visible people in the classroom, so effort is made to diversify instructors and TAs. TA recruitment and selection processes not only emphasize ``clarity, technical proficiency, use of whiteboard, and responsiveness to student questions and needs'' \cite{Kamil2019Gender-balancedBody}, but also cultural competency \cite{Washington2020WhenEnough, Davis2021CulturallyFramework} and readiness to engage with culturally responsive-sustaining CS pedagogy. Classroom and career advice decenters dominant narratives around software engineering jobs in order to highlight opportunities in academic research inside and outside of CS, professional as well non-profit or volunteer opportunities to teach CS in local communities, and roles for computing in social change beyond critique \cite{Abebe2020RolesChange}.

\subsection{Political vision via narrative comparison}

The \emph{critical comparative} approach to ethics and identity orients CCDSA toward a political vision of CS for social justice: design justice \cite{Costanza-Chock2020DesignNeed} and culturally responsive-sustaining pedagogy \cite{Davis2021CulturallyFramework} go a long way to making space for students to engage with sociopolitical values in the computing classroom \cite{Ryoo2020TakeSchools}. Implicit in all of this is a vision of CS for social justice. CCDSA expands on this by explicitly articulating a political vision of CS for social justice through \emph{critical comparison} of narratives and purposes for learning CS2.

From day one of class, teachers emphasize diverse end goals for learning data structures and algorithms: not only the dominant narrative around high-paying jobs in the tech industry, but also marginalized narratives around how data structures and algorithms content knowledge can support research, non-profit, and activist work in revealing social inequities or enabling solutions for local communities. By teaching identity via cultural comparison, teachers break down normative barriers between students and teachers that limit emotional communication---the idea of being distanced and neutral STEM learners is rationalized by the dominant European scientific culture. The establishment of this mutual trust and common understanding between students and teachers enables critique of the ``fraught histories'' \cite{CalabreseBarton2020BeyondLearning} of computing by connecting a thread between the history of CS and present-day inequities. One such thread spans (1) the invention of data structures and algorithms for expressly military purposes, e.g. dynamic programming was invented as a political cover for mathematics research given the uncertainty around post-war US military research funding; into (2) the authoring of computer science as an academic discipline defined by post-war mathematics researchers seeking a new brand for their cognitive mathematics work; finally leading to the (3) the marginalization of many Black women computers who were pushed out of programming once computing became perceived as a profitable field for white men. By engaging this dialogue with students through collaborative teamwork and creative assignments, CCDSA develops students' political vision through understanding data structures and algorithms content knowledge not as a purely ``technical'' and apolitical knowledge, but rather as a knowledge situated in the political vision of the authors of the field.

\section{Conclusion}
\label{sec:conclusion}

The articulation of a political vision for CS centered in social justice requires work not just in a single CS2 course but rather across all of CS education. This is an ambitious project, one that will require us to reprogram the defaults of CS education. But it's also a project that today's CS educators do not need to solve alone: though today's students may be newcomers to CS, they are also precisely the people who will be the authors of our future worlds. When we center ethics, identity, and political vision in our culturally responsive-sustaining pedagogy, we treat students as collaborators and co-creators of education that works for them. Rather than seek assimilation of diverse students into the dominant CS culture, justice-centered approaches ensure the rightful presence of students' political struggles for co-authorship in designing and imagining a new, more just CS education. To create more socially-just worlds, we need to do more than just teach students computer ethics. We need justice-centered approaches to CS in higher education.

\bibliographystyle{ACM-Reference-Format}
\bibliography{ms}

\end{document}